\documentclass[smallcondensed]{svjour3}     
\smartqed  
\usepackage{graphicx}
\begin{document}
\title{The dynamics of entropy uncertainty for qutrit system under the random telegraph noise}



\author{Xiong Xu $^1$        \and
        MaoFa Fang $^1$.
}


\institute{MaofaFang \at
              1.Synergetic Innovation Center for Quantum Effects and Applications, Key Laboratory of Low-dimensional Quantum Structures and Quantum Control of Ministry of Education, School of Physics and Electronics, Hunan Normal University, Changsha, 410081, P.R. China.  \\
              \email{mffang@hunnu.cn}           
           \and
}

\date{Received: date / Accepted: date}
\maketitle
\begin{abstract}
We study the dynamics of quantum memory assist entropy uncertainty for qutrit system coupled to an environment modeled by random matrices. The results show that
the effect of relative coupling strength on entropy uncertainty is opposite in Markov region and non-Markov region, and the influence of a common environment and independent environments in Markov region and non-Markov region is also opposite. One can reduce the entropy uncertainty by
decreasing relative coupling strength or puting the system in two separate environments in the Markov case. In the non-Markov case, the entropy uncertainty can be reduced by increasing the relative coupling strength or by placing the system in a common environment.
\end{abstract}

\section{Introduction}
\label{intro}
The nature of the quantum world is  inherently unpredictable. By far at the most famous statement of unpredictability lies the Heisenberg uncertainty principle about position and momentum, which was first introduced in 1927\cite{Heisenberg.1927}. In fact, for arbitrary observables $R$ and $S$, there is a uncertainty relation which is showed by Robertson in 1929\cite{Robertson.1929},
\begin{equation}
  \Delta R \cdot \Delta \S \geq \frac{|\langle[R,S]\rangle|}{2},
\end{equation}
with the variance $\Delta Q=\sqrt{\langle Q^2\rangle - {\langle Q \rangle}^2}$   (Q is an arbitrary observable).
$\langle \cdot \rangle$ is the expectation of the observable in a quantum system $\rho$, and $[R,S]$ denotes the commutator.
In particular, information theory offers a very versatile, abstract framework that allows us to formalize notions like uncertainty and unpredictability. In 1988, Maassen and Uffink proposed the entropy uncertainty relation based on information entropy\cite{MaassenH.UffinkJ.B.M.:.1988}. It states that
\begin{eqnarray}
   H(R)+H(S)\geq \log_{2}\frac{1}{c},
\end{eqnarray}
where $H(X)$ is Shannon's entropy and c denotes the maximum overlap between any two eigenvectors of the observables R and S.
One of the most important recent developments is the generalization of the uncertainty paradigm, which is the entropy uncertainty relation of quantum memory\cite{RenesJ.M.BoileauJ.C..2008,RenesJ.M.BoileauJ.C..2009}. The quantum memory uncertainty relations can be  understood through a guessing game. It is illustrated  in figure 1, as the following

$(1)$  Bob prepares a bipartite quantum system AB in a state
$\rho_{AB}$. He sends system A to Alice while he keeps system B.

$(2)$  Alice performs one of two possible measurements X
or Z on A and stores the outcome in the classical register K. She communicates her choice to Bob.

$(3)$  Bob’s task is to guess K.
\begin{figure}
 \includegraphics[width=1\textwidth]{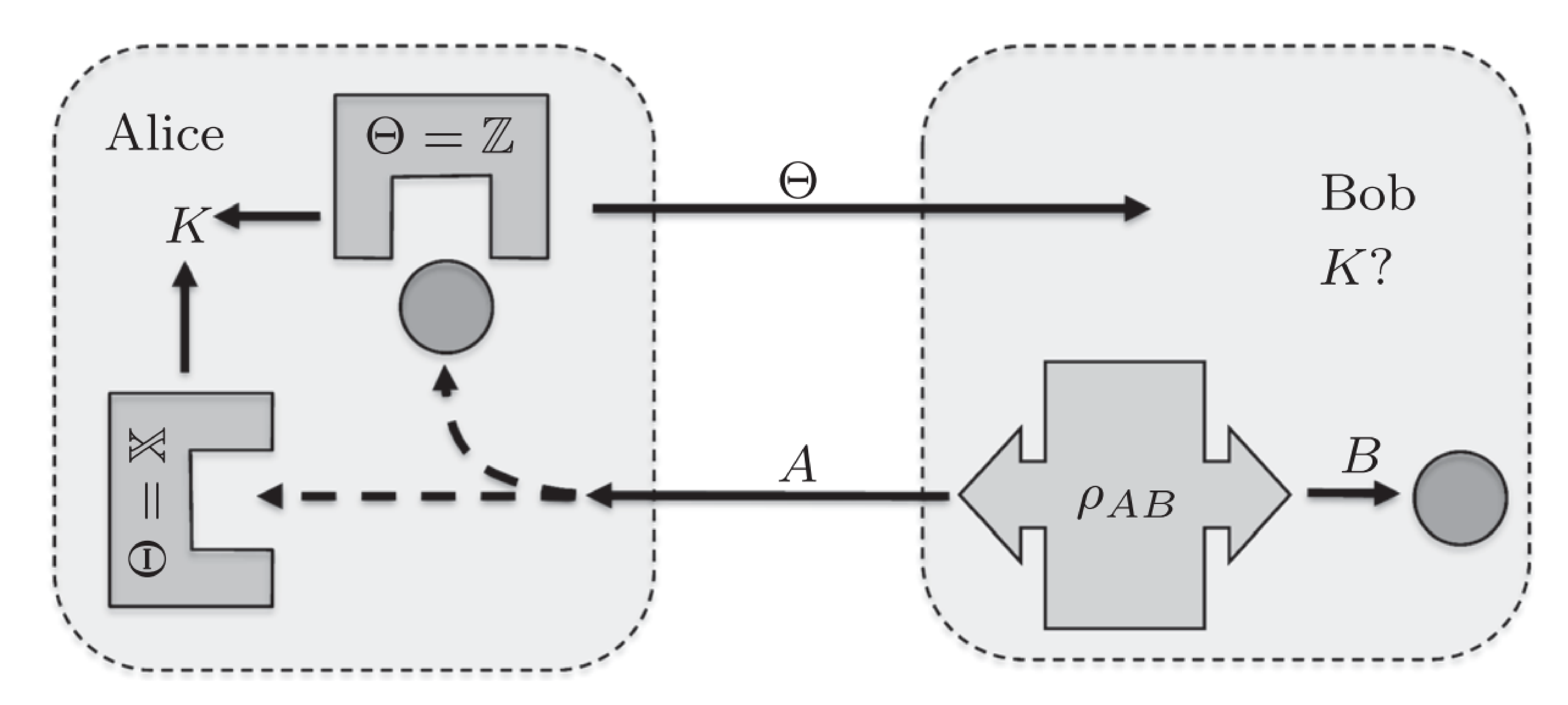}

\caption{The guessing gamen in the presence of a quantum memory system\cite{Berta.2010}. }
\label{fig:1}       
\end{figure}

Indeed, in 2010 Berta et al.\cite{Berta.2010,LiC.F.XuJ.S.XuX.Y.LiK.GuoG.C.:.2011}proved the following entropy uncertainty relation.
\begin{equation}
H(X|B)+H(Z|B)\geq \log_{2}\frac{1}{c}+H(A|B),
\end{equation}
where the conditional entropy $H(X|B)$ is calculated on the classical-quantum state
\begin{equation}
\rho_{XB}=\sum_i(|\psi_i\rangle\langle\psi_i|\otimes I)\rho_{AB}(|\psi_i\rangle\langle\psi_i|\otimes I)
\end{equation}
with $|\psi_i\rangle$ as the eigenstate of the observable $X$ and similarly for $H(Z|B)$.

The uncertainty principle in the presence of memory is important for cryptographic applications\cite{Nataf.2012,Tomamichel.2012,Dupuis.2015} and witnessing entanglement\cite{Zou.2014,Coles.2014,Hall.2012,Prevedel.2011,Hu.2012}. Furthermore, such uncertainty relations are also important for basic physics. For example, interferometry experiments and the quantum-to-classical transition. Recently, there have been many studies on the dynamics of entropy uncertainty in qubit system\cite{Hu.2012,Zhang.2018,Xu.2012,Ming.2018,Huang.2017,JunFeng.2013,LijuanJia.2015,Zheng.2016}, but there are few studies on the dynamics of entropy uncertainty in qutrit system\cite{Guo.2018,YouNengGuo.2018}. Many studies have shown that qutirt system has more advantages than qubit system in quantum information processing\cite{Mair.2001,Fickler.2014,MolinaTerriza.2005,Inoue.2009,Walborn.2006}. For example, key distribution is more secure, quantum locality is much stronger. Therefore, it is meaningful and necessary to study the dynamics of entropy uncertainty in qutrit system.

In this paper, we explore the dynamics of quantum memory assist entropy uncertainty for qutrit system under the random telegraph noise. In Sec. 2, we present our physical model which consider qutrit system under the random telegraph noise. In Sec. 3, the analytic and the graphical result of the entropy uncertainty is investigated in Markovian and non-Markovian regimes.
Finally, we summarize our work in Sec. 4.

\section{Model}
\label{sec:1}
We consider two non-interacting qutirts system which is initially entangled cooupled to an random telegraph noise environment\cite{Carrera.2019,Arthur.2017,Arthur.2018}.  The Hilbert space of the qubit will be labeled
by the subindex $a$ and $b$. We assume that the dynamics in the whole Hilbert space is unitary, the Hamiltonian can be expressed in general as:
\begin{equation}
H=H_{a}\otimes I_{b}+I_{a}\otimes H_{b},
\end{equation}
where $I$ is  the identity matrix of qutrit, $H_{a(b)}$ is the single qutrit Hamiltonian  given by:
\begin{equation}
H_k=\omega _0 I_{k}+\gamma\chi _{k} (t)S_{x}^{k}, k\in \{a,b\}
\end{equation}
The first terms in Eq. (6) represent the free evolution of  qutirt system, and the  second term provides the coupling of random telegraphic noise. $\omega _0$ is the energy frequecy of an isolated qutrit. $g$ is the system-environment coupling constant. $S_{x}^{i}$ is the the spin-1 operator for the x-direction. $\chi _{i} (t)$ stands for the  stochastic variable . The density matrix of the qutrit system for a time t is given by
\begin{equation}
\rho(t)= \langle U(\chi, t)\rho(0) U^\dag(\chi, t)\rangle_{\{\chi\}},
\end{equation}
where $U(\chi, t)$ is the unitary time evolution operator for system, given by
\begin{equation}
 U({\{\chi\}}, t)=U_a(\chi_a, t)\otimes U_b(\chi_b, t),{\{\chi\}}=\{\chi_a,\chi_b \},
\end{equation}
with $U_k(\chi, t)=\exp[-i\int_0^tH_k( \tau) d\tau], k\in \{a,b\}$, assuming $\hbar=1$.
$\rho(0)$ is the initial state, and we consider it as the maximum entangled state
$|\psi\rangle \langle \psi|$ with $|\psi\rangle=\frac{1}{\sqrt{3}}(|00\rangle+|11\rangle+|22\rangle)$. $\langle\cdot \rangle_{\{\chi\}}$ represent the average for a random noisy environment. If both qutrits are coupled to their respective environment called independent environments case. The two-qutrit density matrix is given by the average over the random phase factors for independent environments case
\begin{equation}
 \rho(t)=\langle\langle\rho(\xi_a(t),\xi_b(t))\rangle_{\xi_a}\rangle_{\xi_b}.
\end{equation}
Else two qutrits are coupled to a common environment ($\xi_a=\xi_b$), called a common environment case. The time-evolving state for a common environment case can be given by
\begin{equation}
 \rho(t)=\langle\rho(\xi(t))\rangle_{\xi},
\end{equation}
 with the  random phase factors $\xi_k(t)=-g \int_0^t\chi_k ( \tau) d\tau$. The estimate of the averaged terms of the type $\langle e^{in\xi(t)}\rangle$.
In this work, we focus on the paradigmatic noise:  the random telegraphic noise. For the RTN:
\begin{eqnarray}
\begin{array}{lr}
\langle \cos(n\xi(t))\rangle=\left\{
             \begin{array}{lr}
             e^{-\lambda t}[\cosh(\delta_{n\gamma})+\frac{\lambda}{\delta_{n\gamma}}\sinh(\delta_{n\gamma t)}],\quad&\lambda>n\gamma   \\
\\
              e^{-\lambda t}[\cos(\delta_{n\gamma})+\frac{\lambda}{\delta_{n\gamma}}\sin(\delta_{n\gamma t)}],\quad&\lambda<n\gamma
             \end{array}
\right.
\\
\\
\langle \sin(n\xi(t))\rangle=0,
\end{array}
\end{eqnarray}
where $\delta_{n\gamma}=\sqrt{|\lambda^2-(n\gamma)^2|}$,
with the stochastic parameter $\chi (t)$  describing a fluctuator randomly flipping between the values $\pm1$ at rate $\lambda$.

In all cases, whether it is the independent environment or the common environment, the state evolution has the following form
\begin{eqnarray}
\rho(t)=
\frac{1}{24}\left(
\begin{array}{ccccccccc}
A &    0 & B & 0    & C & 0    & B & 0     & A\\
0 & -2B & 0 & -2B & 0 & -2B & 0 & -2B & 0\\
B & 0    & D & 0    & E & 0    & D & 0    & B\\
0 & -2B & 0 & -2B & 0 & -2B & 0 & -2B & 0\\
C &    0 & E & 0    & F & 0    & E & 0     & C\\
0 & -2B & 0 & -2B & 0 & -2B & 0 & -2B & 0\\
B & 0    & D & 0    & E & 0    & D & 0    & B\\
0 & -2B & 0 & -2B & 0 & -2B & 0 & -2B & 0\\
A &    0 & B & 0    & C & 0    & B & 0     & A
\end{array}
\right).
\end{eqnarray}
For the  independent environments case,
\begin{eqnarray}
\begin{array}{ll}
A=3+4\langle e^{i\xi_a(t)}\rangle\langle e^{i\xi_b(t)}\rangle+\langle e^{i2\xi_a(t)}\rangle\langle e^{i2\xi_b(t)}\rangle,\\
\\
B=-1+\langle e^{i2\xi_a(t)}\rangle\langle e^{i2\xi_b(t)}\rangle,\\
\\
C=2(1+2\langle e^{i\xi_a(t)}\rangle\langle e^{i\xi_b(t)}\rangle+\langle e^{i2\xi_a(t)}\rangle\langle e^{i2\xi_b(t)}\rangle),\\
\\
D=3-4\langle e^{i\xi_a(t)}\rangle\langle e^{i\xi_b(t)}\rangle+\langle e^{i2\xi_a(t)}\rangle\langle e^{i2\xi_b(t)}\rangle,\\
\\
E=2(1-2\langle e^{i\xi_a(t)}\rangle\langle e^{i\xi_b(t)}\rangle+\langle e^{i2\xi_a(t)}\rangle\langle e^{i2\xi_b(t)}\rangle),\\
\\
F=8+2\langle e^{i\xi_a(t)}\rangle\langle e^{i\xi_b(t)}\rangle.
\end{array}
\end{eqnarray}
For  the common environment case,
\begin{eqnarray}
\begin{array}{lllll}
A=3+4\langle e^{i2\xi(t)}\rangle +\langle e^{i4\xi(t)}\rangle,\\
\\
B=-1+\langle e^{i4\xi(t)}\rangle,\\
\\
C=2(1+2\langle e^{i2\xi(t)}\rangle+\langle e^{i4\xi(t)}\rangle),\\
\\
D=3-4\langle e^{i2\xi(t)}\rangle+\langle e^{i4\xi(t)}\rangle,\\
\\
E=2(1-2\langle e^{i2\xi(t)}\rangle+\langle e^{i4\xi(t)}\rangle),\\
\\
F=8+2\langle e^{i2\xi(t)}\rangle.
\end{array}
\end{eqnarray}
\section{Results and Discussion}
In this section, we present the analytic and the graphical results of the entropy uncertainty of the qutrit system under Random telegraphic noise. In the previous section, we have given the formal solution of time-evolving state so we can get the conditional entropy after qutrit A was measured by Alice($S_x$ or $S_z$).
\begin{eqnarray}
H(S_x|B)&=&H(\rho_{S_xB})-H(\rho_B),\nonumber
\\
H(S_z|B)&=&H(\rho_{S_zB})-H(\rho_B),
\end{eqnarray}
where $H(\rho)$ is the von-Neumann entropy by $H(\rho)=-\sum_i\lambda_i\log_2\lambda_i$ and $\rho_B=Tr_A(\rho(t))$. The analytic results are as follows:

$(1)$ eigenvalues of $\rho_{S_xB}$: $\lambda^x_1=\lambda^x_2=\lambda^x_3=\frac{1}{3}$, and others are zero.

$(2)$ eigenvalues of $\rho_{S_zB}$:
\begin{eqnarray}
\lambda^z_1&=&0\nonumber\\
\lambda^z_2&=&\frac{1-\beta }{6}\nonumber\\
\lambda^z_3&=&\frac{1+\beta }{6}\nonumber\\
\lambda^z_4&=&\lambda^z_5=\frac{1-\beta }{12}\nonumber\\
\lambda^z_6&=&\lambda^z_7=\frac{1}{24} \left(-\sqrt{16 \alpha ^2+(\beta -1)^2}+\beta +3\right)\nonumber\\
\lambda^z_8&=&\lambda^z_9=\frac{1}{24} \left(\sqrt{16 \alpha ^2+(\beta -1)^2}+\beta +3\right)
\end{eqnarray}

$(3)$ eigenvalues of $\rho_{B}$: $\lambda^x_1=\lambda^x_2=\lambda^x_3=\frac{1}{3}$.
where $\alpha=\langle e^{i\xi_a(t)}\rangle\langle e^{i\xi_b(t)}\rangle$, $\beta=\langle e^{i2\xi_a(t)}\rangle\langle e^{i2\xi_b(t)}\rangle$ for the  independent environments case, $\alpha=\langle e^{i2\xi(t)}\rangle$, $\beta=\langle e^{i4\xi(t)}\rangle$ for the common environments case. In this paper we assume $\xi_a(t)=\xi_b(t)$ and  introduce the relative interaction strengths $g=\gamma/\lambda$ in RTN case.
Therefore, the entropy uncertainty can be calculated as following:
\begin{eqnarray}
U_L=-\sum_i\lambda_i^z\log_2\lambda_i^z+\log_23.
\end{eqnarray}

\subsection{The Markovian regime}

\begin{figure}[htpb]
\begin{center}
 \includegraphics[width=0.45\textwidth]{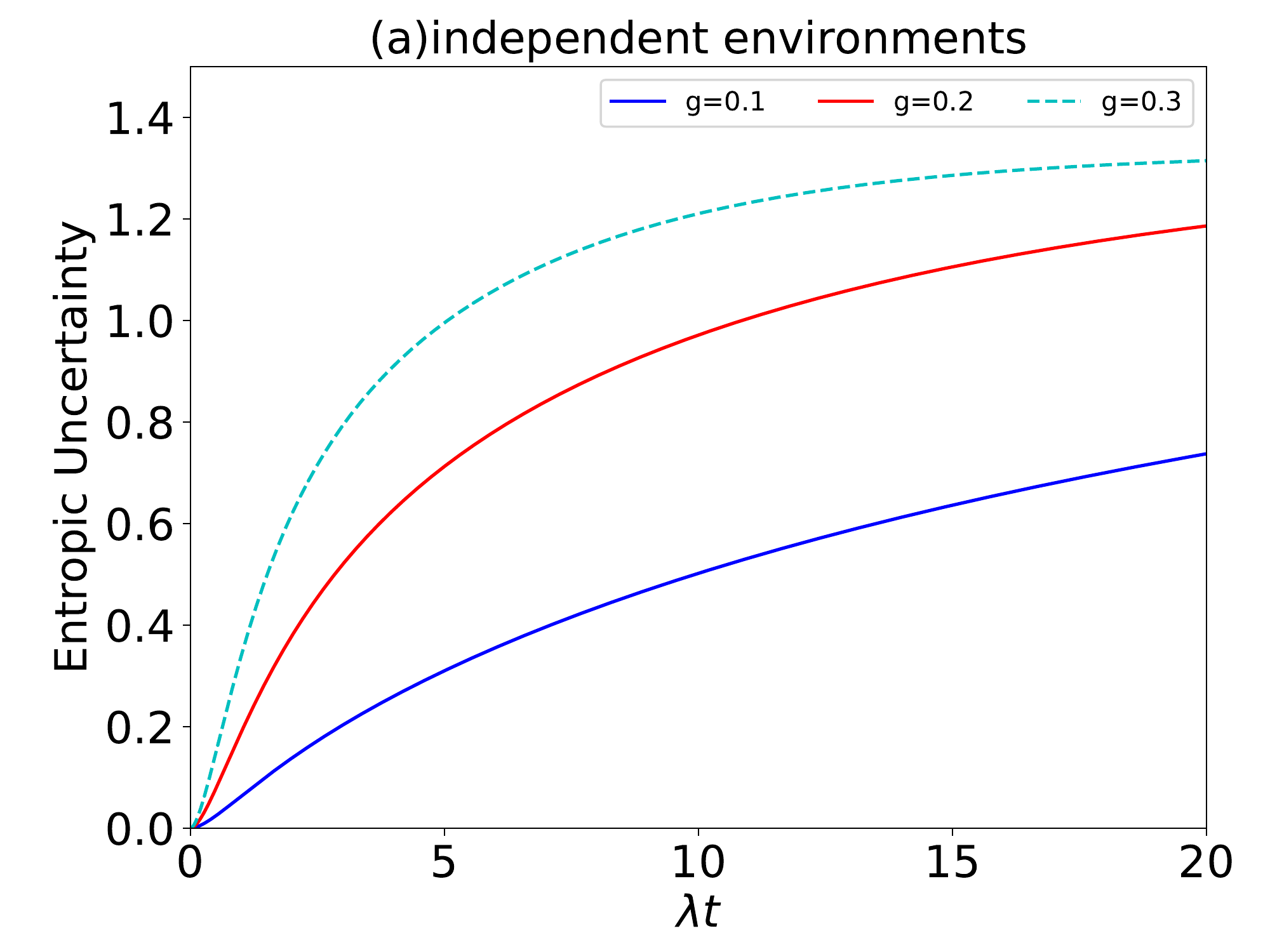}
 \includegraphics[width=0.45\textwidth]{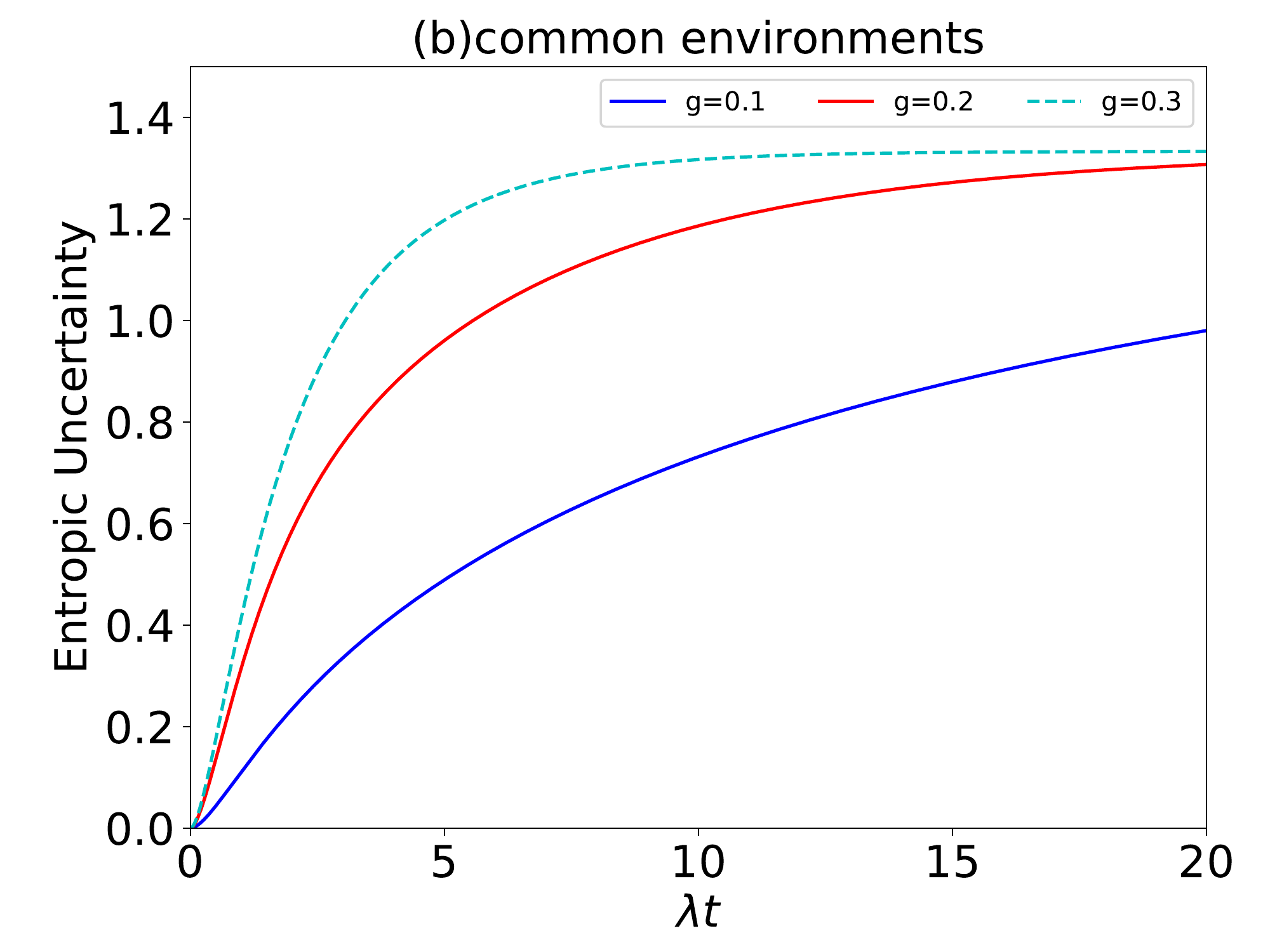}
\caption{The evolution of the entropy uncertaity under the Markov RTN both in  the  independent environments case $(a)$ and the common environments case$(b)$ for fixed values of relative interaction strengths g. }
\label{fig:2}       
\end{center}
\end{figure}

In the Markovian regime, the entropy uncertainty keep increasing with time both in the  independent environments case and the common environments case which are shown in the Fig.\ref{fig:2}. One can see that the entropy uncertainty grows faster in the common environment case than in the independent environments case when the relative interaction strength g is the same. In the independen environments case, the increase of entropy uncertainty is accelerated with the increase of relative interaction strength g, and the higher the relative coupling strength is, the greater the entropy uncertainty. The same is true in the common environment case.

\subsection{The Non-Markovian regime}
\begin{figure}[htpb]
\begin{center}
 \includegraphics[width=0.45\textwidth]{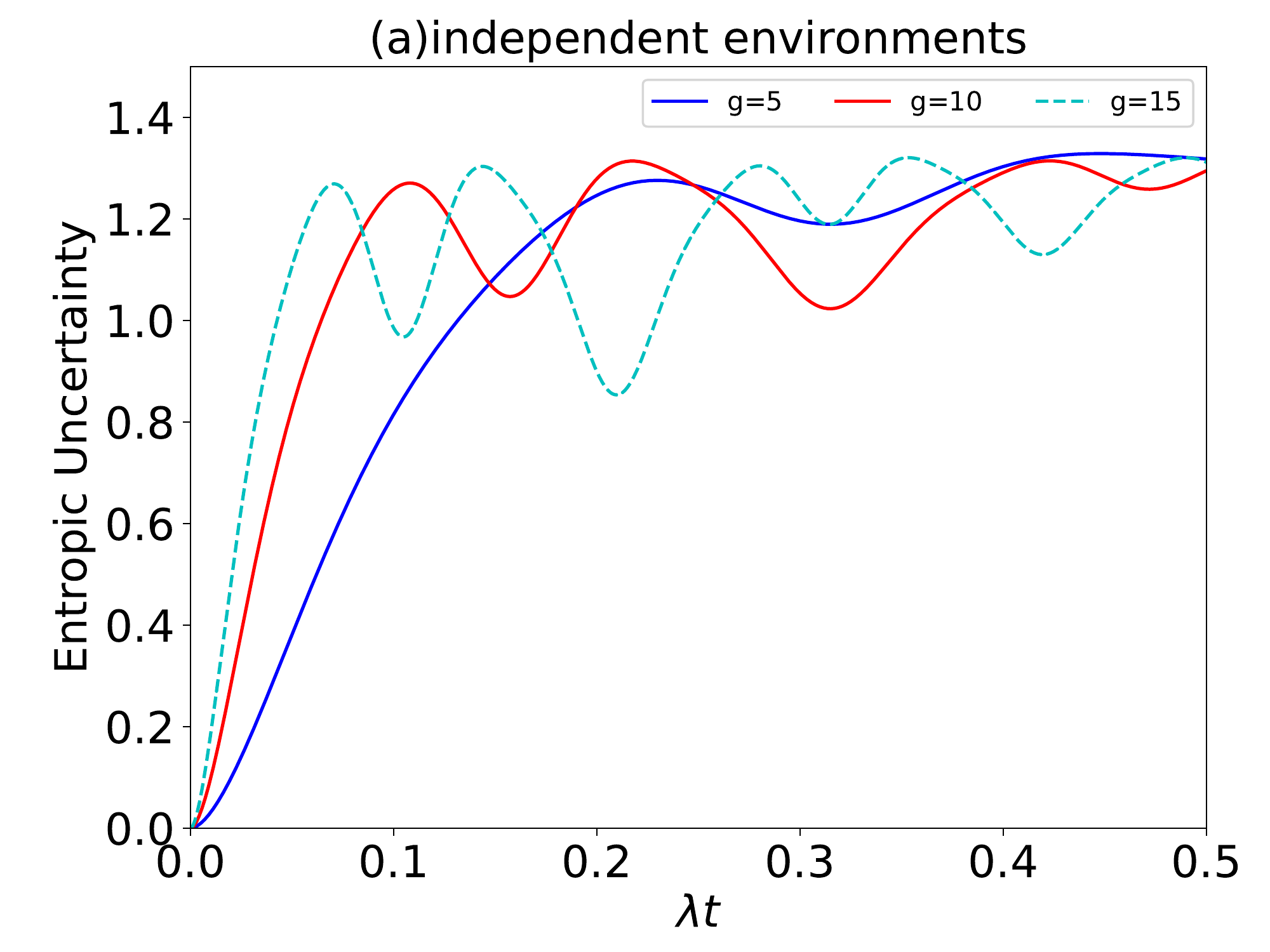}
 \includegraphics[width=0.45\textwidth]{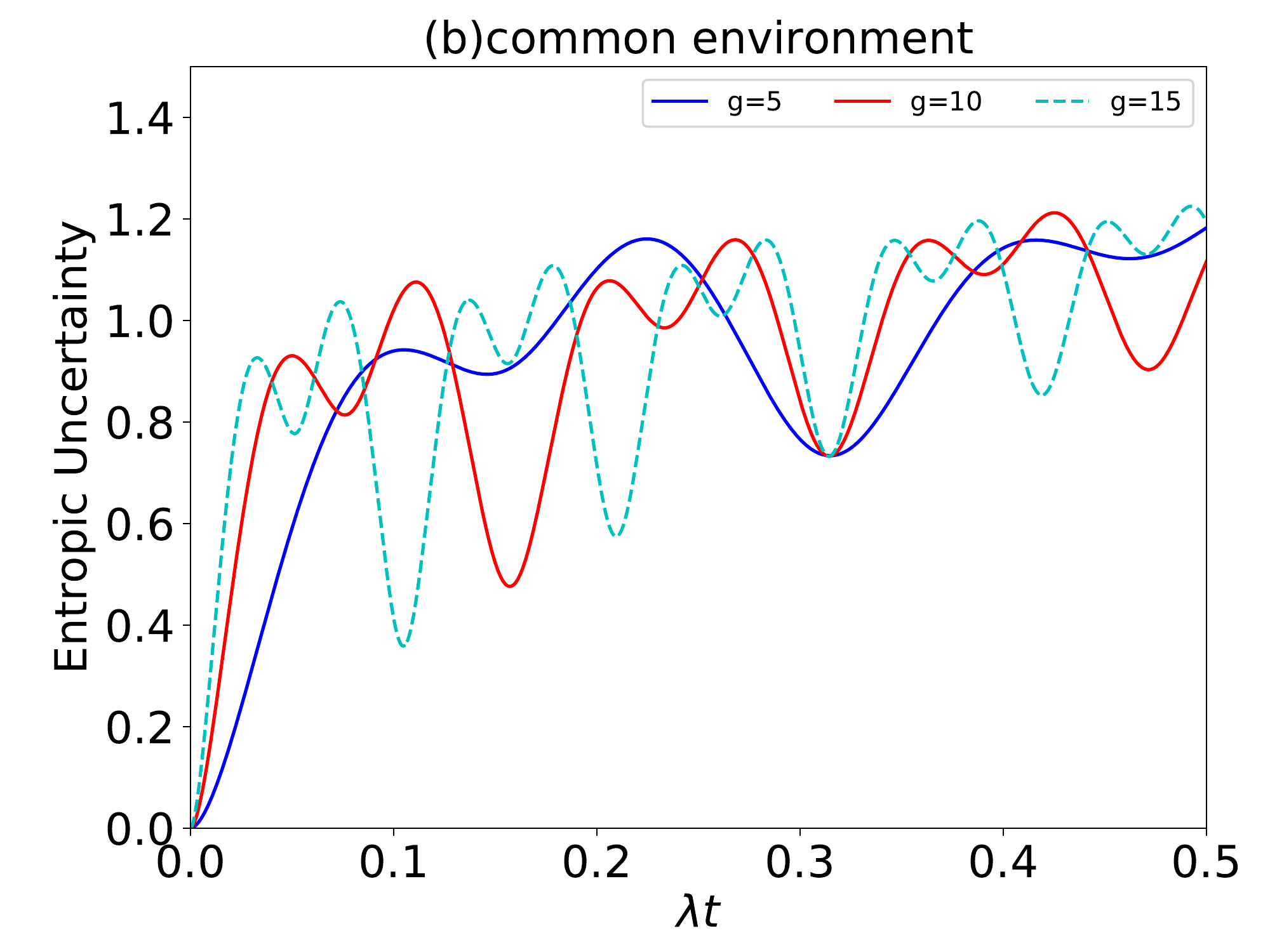}
\caption{The evolution of the entropy uncertaity under the Non-Markovian RTN both in  the  independent environments case $(a)$ and the common environments case$(b)$ for fixed values of relative interaction strength g. }
\label{fig:3}       
\end{center}
\end{figure}
As expected from the non-Markovian nature of the noise, the entropy uncertaity is oscillating function of time shown in Fig.\ref{fig:3}. Oscillations of the entropy uncertaity become more and more prominent in the qutrit system as the relative strength g increases. Contrary to Markov's case, it is clearly that as the interaction intensity increases, the minimum value of entropy uncertainty decreases both in the independent environments case and the common environment case. In the  non-Markovian regime, the entropy uncertainty in the common environment case is smaller than that in the independent environment case which is also different from the Markovian regime.

\section{Conclusions}

To sum, we have investigated the dynamics of quantum memory assist the entropy uncertainty for a qutrit system under the random telegraph noise (RTN). We have found that in the Markovian regime, two separate environments can help reduce entropy uncertainty,which can also be reduced by decreasing the relative coupling strength g. In the Non-Markovian regime, a common environment is more favorable to keep the entropy uncertainty at a low value. And in this case, the relative coupling strength g plays an opposite role, increasing g can reduce the entropy uncertainty. These results enable us to select different operations for different conditions to reduce entropy uncertainty when dealing with quantum information problems.

\section*{Acknowledgments}
This work is supported by the National Natural Science Foundation of China (Grant No.11374096).
\bibliographystyle{unsrt}
\bibliography{paper3}

\begin{thebibliography}{10}

\bibitem{Heisenberg.1927}
W.~Heisenberg.
\newblock The actual content of quantum theoretical kinematics and mechanics.
\newblock {\em Z. Phys.}, 43:172, 1927.

\bibitem{Robertson.1929}
H.~P. Robertson.
\newblock The uncertainty principle.
\newblock {\em Phys. Rev.}, 34:163, 1929.

\bibitem{MaassenH.UffinkJ.B.M.:.1988}
H.~Maassen and J.B.M. Uffink.
\newblock Generalized entropic uncer- tainty relations.
\newblock {\em Phys. Rev. Lett.}, 60:1103, 1988.

\bibitem{RenesJ.M.BoileauJ.C..2008}
J.M. Renes and J.C. Boileau.
\newblock Physical underpinnings of privacy.
\newblock {\em Phys. Rev. A}, 78:032335, 2008.

\bibitem{RenesJ.M.BoileauJ.C..2009}
J.M. Renes and J.C. Boileau.
\newblock Conjectured strong complementary information tradeoff.
\newblock {\em Phys. Rev. Lett.}, 103:020402, 2009.

\bibitem{Berta.2010}
Mario Berta, Matthias Christandl, Roger Colbeck, Joseph~M. Renes, and Renato
  Renner.
\newblock The uncertainty principle in the presence of quantum memory.
\newblock {\em Nature Phys (Nature Physics)}, 6(9):659--662, 2010.

\bibitem{LiC.F.XuJ.S.XuX.Y.LiK.GuoG.C.:.2011}
C.F. Li, J.S. Xu, X.Y. Xu, K.~Li, and G.C. Guo.
\newblock Experimental investigation of the entanglement-assisted entropic
  uncertainty principle.
\newblock {\em Nat. Phys.}, 7:752, 2011.

\bibitem{Nataf.2012}
P.~Nataf, M.~Dogan, and K.~L. Hur.
\newblock Heisenberg uncertainty principle as a probe of entanglement entropy:
  Application to superradiant quantum phase transitions.
\newblock {\em Phys. Rev. A}, 86:043807, 2012.

\bibitem{Tomamichel.2012}
M.~Tomamichel, C.~CW. Lim, N.~Gisin, and R.~Renner.
\newblock Tight finite-key analysis for quantum cryptography.
\newblock {\em Nature Commun.}, 3:634, 2012.

\bibitem{Dupuis.2015}
F.~Dupuis, O.~Fawzi, and S.~Wehner.
\newblock Entanglement sampling and applications.
\newblock {\em IEEE Trans. Inf. Theory}, 61:1093, 2015.

\bibitem{Zou.2014}
Hong-Mei Zou, M.~F. Fang, Bai-Yuan Yang, You-neng Guo, Wei He, and Shi-Yang
  Zhang.
\newblock The quantum entropic uncertainty relation and entanglement witness in
  the two-atom system coupling with the non-markovian environments.
\newblock {\em Physica Scripta}, 89(11):115101, 2014.

\bibitem{Coles.2014}
P.~J. Coles and M.~Piani.
\newblock Complementary sequential measurements generate entanglement.
\newblock {\em Phys. Rev. A}, 89:010302, 2014.

\bibitem{Hall.2012}
M.J.W. Hall and H.M. Wiseman.
\newblock Heisenberg-style bounds for arbitrary estimates of shift parameters
  including prior information.
\newblock {\em New J. Phys.}, 14:033040, 2012.

\bibitem{Prevedel.2011}
R.~Prevedel, D.~R. Hamel, R.~Colbeck, K.~Fisher, and K.~J. Resch.
\newblock Experimental investigation of the uncertainty principle in the
  presence of quantum memory.
\newblock {\em Nat. Phys.}, 7:757, 2011.

\bibitem{Hu.2012}
M.~Hu and H.~Fan.
\newblock Quantum-memory-assisted entropic uncertainty principle,
  teleportation, and entanglement witness in structured reservoirs.
\newblock {\em Phys. Rev. A}, 86:032338, 2012.

\bibitem{Zhang.2018}
Yanliang Zhang, Maofa Fang, Guodong Kang, and Qingping Zhou.
\newblock Controlling quantum memory-assisted entropic uncertainty in
  non-markovian environments.
\newblock {\em Quantum Information Processing}, 17(3):172, 2018.

\bibitem{Xu.2012}
Z.~Y. Xu, W.~L. Yang, and M.~Feng.
\newblock Quantum-memory-assisted entropic uncertainty relation under noise.
\newblock {\em Phys. Rev. A}, 86:012113, 2012.

\bibitem{Ming.2018}
Fei Ming, Dong Wang, Wei-Nan Shi, Ai-Jun Huang, Wen-Yang Sun, and Liu Ye.
\newblock Entropic uncertainty relations in the heisenberg xxz model and its
  controlling via filtering operations.
\newblock {\em Quantum Information Processing}, 17(4):89, 2018.

\bibitem{Huang.2017}
Ai-Jun Huang, Dong Wang, Jia-Ming Wang, Jia-Dong Shi, Wen-Yang Sun, and Liu Ye.
\newblock Exploring entropic uncertainty relation in the heisenberg xx model
  with inhomogeneous magnetic field.
\newblock {\em Quantum Information Processing}, 16(8):204, 2017.

\bibitem{JunFeng.2013}
{Jun Feng}, {Yao-Zhong Zhang}, {Mark D. Gould}, and {Heng Fan}.
\newblock Entropic uncertainty relations under the relativistic motion.
\newblock {\em Physics Letters B}, 726(1):527--532, 2013.

\bibitem{LijuanJia.2015}
{Lijuan Jia}, {Zehua Tian}, and {Jiliang Jing}.
\newblock Entropic uncertainty relation in de sitter space.
\newblock {\em Annals of Physics}, 353:37--47, 2015.

\bibitem{Zheng.2016}
Xiao Zheng and Guo-Feng Zhang.
\newblock The effects of mixedness and entanglement on the properties of the
  entropic uncertainty in heisenberg model with dzyaloshinski--moriya
  interaction.
\newblock {\em Quantum Information Processing}, 16(1):1, 2016.

\bibitem{Guo.2018}
You-neng Guo, Mao-fa Fang, and Ke~Zeng.
\newblock Entropic uncertainty relation in a two-qutrit system with external
  magnetic field and dzyaloshinskii--moriya interaction under intrinsic
  decoherence.
\newblock {\em Quantum Information Processing}, 17(7):187, 2018.

\bibitem{YouNengGuo.2018}
{You-Neng Guo}, {Mao-Fa Fang}, {Qing-Long Tian}, {Zheng-Da Li}, and {Ke Zeng}.
\newblock Exploration of the entropic uncertainty relation for a qutrit system
  under decoherence.
\newblock {\em Laser Physics Letters}, 15(10):105205, 2018.

\bibitem{Mair.2001}
Alois Mair, Alipasha Vaziri, Gregor Weihs, and Anton Zeilinger.
\newblock Entanglement of the orbital angular momentum states of photons.
\newblock {\em Nature}, 412(6844):313--316, 2001.

\bibitem{Fickler.2014}
Robert Fickler, Radek Lapkiewicz, Marcus Huber, Martin~P.J. Lavery, Miles~J.
  Padgett, and Anton Zeilinger.
\newblock Interface between path and orbital angular momentum entanglement for
  high-dimensional photonic quantum information.
\newblock {\em Nature Communications}, 5(1):4502, 2014.

\bibitem{MolinaTerriza.2005}
G.~Molina-Terriza, A.~Vaziri, R.~Ursin, and A.~Zeilinger.
\newblock Experimental quantum coin tossing.
\newblock {\em Phys. Rev. Lett.}, 94(4):040501, 2005.

\bibitem{Inoue.2009}
R.~Inoue, T.~Yonehara, Y.~Miyamoto, M.~Koashi, and M.~Kozuma.
\newblock Measuring qutrit-qutrit entanglement of orbital angular momentum
  states of an atomic ensemble and a photon.
\newblock {\em Phys. Rev. Lett.}, 103(11):110503, 2009.

\bibitem{Walborn.2006}
S.~P. Walborn, D.~S. Lemelle, M.~P. Almeida, and P.~H.~Souto Ribeiro.
\newblock Quantum key distribution with higher-order alphabets using spatially
  encoded qudits.
\newblock {\em Phys. Rev. Lett.}, 96(9):090501, 2006.

\bibitem{Carrera.2019}
M.~Carrera, T.~Gorin, and C.~Pineda.
\newblock Markovian and non-markovian dynamics induced by a generic
  environment.
\newblock {\em Phys. Rev. A}, 100(4):042322, 2019.

\bibitem{Arthur.2017}
Tsamouo~Tsokeng Arthur, Tchoffo Martin, and Lukong~Cornelius Fai.
\newblock Quantum correlations and coherence dynamics in qutrit--qutrit systems
  under mixed classical environmental noises.
\newblock {\em International Journal of Quantum Information}, 15(06):1750047,
  2017.

\bibitem{Arthur.2018}
Tsamouo~Tsokeng Arthur, Tchoffo Martin, and Lukong~Cornelius Fai.
\newblock Disentanglement and quantum states transitions dynamics in
  spin-qutrit systems: dephasing random telegraph noise and the relevance of
  the initial state.
\newblock {\em Quantum Information Processing}, 17(2):37, 2018.

\end{thebibliography}
%




\end{document}